\documentclass[fleqn,10pt]{wlscirep}
\usepackage[utf8]{inputenc}
\usepackage[T1]{fontenc}
\usepackage{textcomp}
\usepackage{hyperref}

\usepackage{cleveref}
\usepackage{multirow}

\title{High-throughput Screening of the Mechanical Properties of Peptide Assemblies}

\author[1,*]{Sarah K. Yorke}
\author[2,3,*]{Zhenze Yang}
\author[1]{Aviad Levin}
\author[4]{Alice Ray}
\author[4]{Jeremy Owusu Boamah}
\author[1,4,$\dag$]{Tuomas P. J. Knowles}
\author[2,5,6,$\ddag$]{Markus J. Buehler}
\affil[1]{Yusuf Hamied Department of Chemistry, University of Cambridge, Lensfield Road, Cambridge, CB2 1EW, UK}
\affil[2]{Laboratory for Atomistic and Molecular Mechanics, Department of Civil and Environmental Engineering, Massachusetts Institute of Technology, 77 Massachusetts Ave. Room 1-235A-B, Cambridge, 02139, MA, USA}
\affil[3]{Department of Materials Science and Engineering, Massachusetts Institute of Technology, 77 Massachusetts Ave., Cambridge, 02139, MA, USA}
\affil[4]{Cavendish Laboratory, University of Cambridge, J J Thomson Avenue, Cambridge, CB3 0HE, UK}
\affil[5]{Center for Computational Engineering, Massachusetts Institute of Technology, 77 Massachusetts Ave., Cambridge, 02139, MA, USA}
\affil[6]{Center for Materials Science and Engineering, Massachusetts Institute of Technology, 77 Massachusetts Ave., Cambridge, 02139, MA, USA}
\affil[*]{These authors contribute equally.}
\affil[$\dag$]{tpjk2@cam.ac.uk}
\affil[$\ddag$]{mbuehler@mit.edu}

\begin{abstract}
\noindent Peptides are recognized for their varied self-assembly behaviors, forming a wide array of structures and geometries, such as spheres, fibers, and hydrogels, each presenting a unique set of material properties. The functionalities of these materials hold expectional interest for applications in biology, medicine, photonics, nanotechnology and the food industry. In specific, the ability to exploit peptides as viable and sustainable mechanical materials requires sequence design that enables superior performance, notably a high Young's modulus. As the peptide sequence space is vast, however, even a slight increase in sequence length leads to an exponential increase in the number of potential peptide sequences to be characterized. Here, we combine coarse-grained molecular dynamics simulations, atomic force microscopy experiments and machine learning models to correlate the sequence length and composition with the mechanical properties of self-assembled peptides. We calculate the Young's modulus for all possible amino acid sequences of di- and tripeptides using high-throughput coarse-grained methods, and validate these calculations through in-situ mechanical characterization. For pentapeptides, we select and calculate properties for a subset of sequences to train a machine learning model, which allows us to predict the modulus for other sequences. The combined workflow not only identifies promising peptide candidates with exceptional mechanical performances, but also extends current understanding of the sequence-to-function relationships for peptide materials, for specific applications.
\end{abstract}
\begin{document}

\flushbottom
\maketitle
%
%
\thispagestyle{empty}

\section*{Introduction}
Nature utilises polypeptides to form a wide range of self-assembled structures with unique mechanical properties. The silk in a spider's web \cite{Cranford2012, Nova2010, 10.1063/5.0097589} and collagen in skin \cite{doi:10.1073/pnas.0603216103, doi:10.1021/nl103943u} are some of the strongest known natural materials. In drawing inspiration from such systems, many strong, sustainable and biodegradable materials have been developed in recent years. Key examples of these include soy based transparent films used for packaging applications,\cite{RN308,GUERRERO2010145} and silk-based microgels which can be used as biomedical materials.\cite{B916319K,D3NH00385J, polym11121933} However, utilizing naturally extracted proteins is limited by their complexity, meaning that modulating their mechanical properties and realising highly specialized applications, such as piezoelectric generators and tissue scaffolds, can be difficult to obtain. An alternative approach to expanding on the use of bio-derived proteins is to use short peptides consisting of only a few amino acid building blocks. These offer highly tunable chemical structures due to their short length, but maintain the bio-compatibility and biodegradababilty of biomolecules. \\
\\
As a consequence, short peptide self-assembly has been explored and expanded in the past two decades. Early work showed the formation of fibres from diphenylalanine (FF), and the assembly of nanotubular structures from cyclic octopeptides.\cite{Reches2003,RN304} Since this pioneering work, a wide variety of materials have been fabricated from short peptides, including rigid hydrogels formed from Fmoc-diphenylalanine (Fmoc-FF) \cite{https://doi.org/10.1002/adma.200501522, https://doi.org/10.1002/adma.200501765, C9TB01043B}, which is based upon a rigid interlocking $\beta$-sheet structure.\cite{https://doi.org/10.1002/adma.200701221} Hydrogels assembled from short peptide building blocks are of particular interest due to their potential use in biomedical applications, such as cell scaffolds for tissue culture\cite{RESTU2020110746, gelain_self-assembling_2021} and drug delivery vehicles.\cite{10.3389/fchem.2021.770102, gels8110706} Short peptides have also been manipulated to form nanotublar structures, through the design of amphiphilic and surfactant like structures.\cite{D0CC04299D, doi:10.1073/pnas.1730609100} \\
\\
While a range of peptides have been investigated thus far, as the sequence space increases dramatically with the length of peptides, it is challenging to conduct systematic experimental scans of the self-assembly propensity and associated properties on extended peptide sequences. A polypeptide with N amino acids could have $20^N$ possible sequences, leading to millions of combinations that all need synthesising, and hence scanning the sequence space instead lends itself to dry-lab approaches. This problem was first addressed in 2011, in which coarse-grained (CG) molecular dynamics (MD) simulations were conducted for each dipeptide to predict their aggregation propensity (AP).\cite{doi:10.1021/jz2010573} Extending on this approach, both atomistic and CG MD simulations have been employed to predict tripeptide self-assembly.\cite{RN305, 10.1371/journal.pcbi.1003718} However, these methods are computationally expensive, esepcailly as sequence length increases. More recently, machine learning (ML) and human-in-the-loop approaches have instead been adopted in order combine the high-throughput of computational methods and human knowledge to predict the performance of self-assembling peptides and hydrogelators specifically.\cite{RN307, RN306}\\
\\
Most prior work focuses on peptide and protein self-assembly propensity specifically, but does not correlate aggregation propensity with mechanical properties. Mechanical performances are vital for polypeptides to maintain their functional integrity in diverse applications. For example, a film used for food packaging must have a Young's modulus and tensile strength comparable to that of low density polyethylene (173 and 7.2 MPa respectively). Conversely, artificial silk must have a comparable tensile strength to spider dragline silk (0-8-1.5 GPa)\cite{WANG20224396, article_rising} in order to be a suitable and sustainable alternative to Kevlar. Experimental measurements such as Atomic Force Microscopy (AFM) \cite{Stolz2004, PhysRevLett.91.098101} and computational approaches \cite{C9NA00621D, doi:10.1021/acs.macromol.0c00138} have been widely employed to investigate the mechanical stability of polypeptide assemblies. Intermolecular interactions and secondary structures significantly influence the material's stiffness and deformation mechanisms,\cite{RN309} Hence, by combining these approaches we can further elucidate the relationship between peptide sequence, assembly conditions and Young's modulus, and gain mechanistic insight into the formation and evolution of structural proteins in biology. This can, in turn, provide a framework for the design of stiff and durable biomaterials.\\
\\
In this work, we address this gap through coupled computational methods with extensive experimental validation to measure the mechanical properties for peptides of various lengths; dipeptides, tripeptides and pentapeptides. These lengths allow us to compare 2 computational approaches, a full MD approach and a coupled MD-ML for the 3.2million pentapeptide sequences. We develop a high-throughput workflow based on CGMD simulations to efficiently assemble peptide molecules and assess their mechanical behaviors with a high degree of automation. The accuracy of the computational approach is validated by experimental mechanical characterization of a range of di- and tripeptides at the nanoscale, using AFM nanomechanical mapping. We exhaustively calculate the Young's modulus for all possible self-assembling amino acid sequences of di- and tri-peptides, and subsequently identify the peptides with the highest modulus values. In the case of pentapeptides, given the millions of possible sequences, we train an ML model to predict the modulus of given peptide sequences and collect thousands of data points collected using our high-throughput workflow. This approach is further validated using imaging and nanomechanical mapping of a selection of pentapeptides. Overall, we find that the modulus values calculated using MD for di- and tri-peptides are an accurate indicator for the relative modulus of our materials.

\section*{Results}
\subsection*{Overall workflow}
\begin{figure}
  \includegraphics[width=\linewidth]{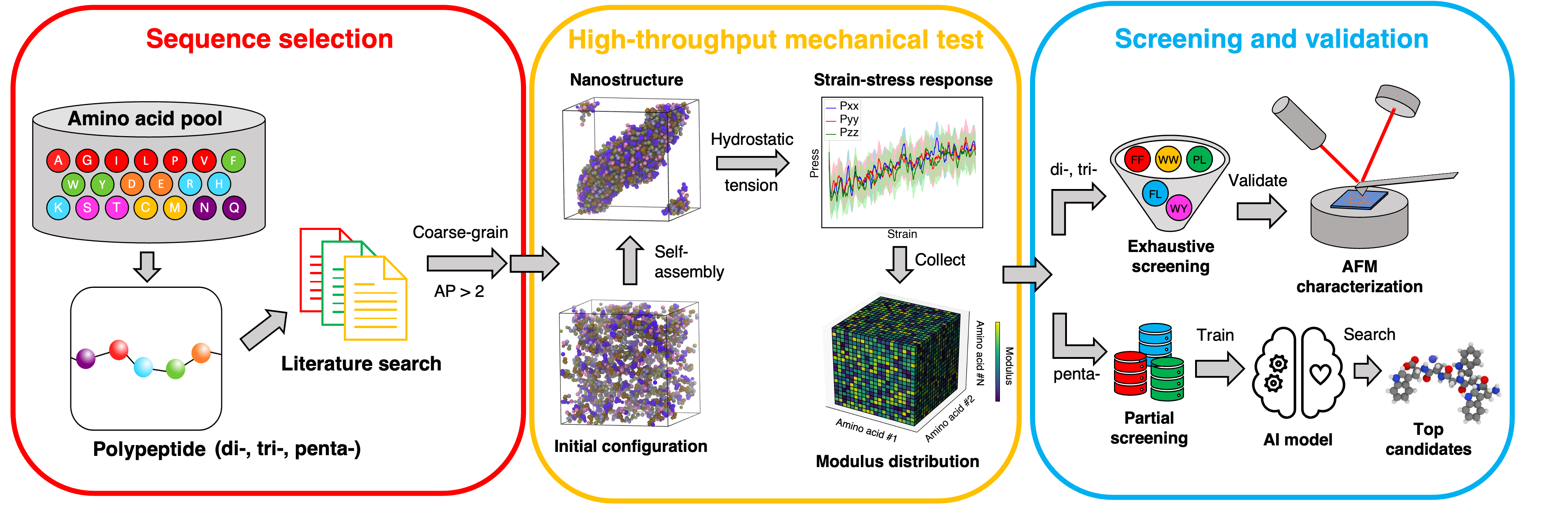}
  \caption{\textbf{Full workflow.} The workflow used in this work to investigate the mechanical properties of polypeptide self-assemblies can be divided into three major steps: 1) \textbf{Sequence selection}: Based on the number of amino acids in the peptide sequence, we investigate di-, tri- and pentapeptides. We first screen these peptides to collect sequences that can form aggregates for mechanical testing with results from previous studies in the literature. 2) \textbf{High-throughput mechanical test}: With the selected sequences, the polypeptides are converted into CG representation based on MARTINI force field.\cite{martini} CG MD simulation is employed to assemble the polypeptides, followed by hydrostatic tension test using NEMD. The Young's moduli of different peptides are derived from strain-stress responses and then collected as datasets. 3) \textbf{Screening and validation}: With the datasets curated by high-throughput CG MD simulations, we can screen the peptides based on their Young's modulus. For di(tri)peptide, we exhaustively screen all possible sequences and validate the simulation results with experimental characterization using AFM. For longer peptides like pentapeptide, we select a proportion of sequences in the design space for step 2) and train an AI-based surrogate model with the data. With the ML model, we can predict modulus for the rest of sequences and search for optimal candidates.}
  \label{fig:1}
\end{figure}

The overall workflow has three key steps (Fig. \ref{fig:1}): 1) Sequence selection, 2) High-throughput mechanical testing and 3) Screening and validation. In the sequence selection step, we collect peptide sequences that are known self-assemble into nanostructures, defined as having an aggregation propensity above 2. The self-assembly behaviors of short peptides has been a focus of various previous studies,\cite{doi:10.1021/jz2010573, RN305, Batra2022} from which we gather the information to create the peptide sequence dataset for modulus investigation. The short peptides are divided based on sequence length into di-, tri- and penta-peptides. \\
\\
Once the self-assembling sequences are compiled, we leverage CG MD simulations to investigate the mechanical properties of these materials using high-throughput mechanical test. We first generate the atomistic structures of peptide monomers and convert them into CG representation using MARTINI force field.\cite{martini} Hundreds of short peptide monomers are inserted into a cubic simulation box and solvated with water. The high-throughput mechanical test comprises of two stages: initially, the peptide system undergoes energy minimization and relaxation from a random initial configuration to allow self-assembly; subsequently, a hydrostatic tension test is conducted by expanding the simulation box along all three dimensions. Based on the strain-stress response of the peptides during the hydrostatic tension test, we can then calculate the Young's moduli of the assembled nanostructures.\\
\\
At the screening and validation stage, for dipeptides and tripeptides, given their relatively limited sequence space, we exhaustively calculate the moduli for all sequences capable of self-assembling. In contrast, for the longer pentapeptides, the vast number of possible sequences ($\sim$ millions) means our computational capacity is only sufficient to determine the moduli of a proportion of these sequences (partial screening). To evaluate the modulus of the remaining sequences, we train a machine learning (ML) model using these calculated sequences. The model is then implemented to predict the modulus for all other pentapeptides, allowing us to identify the most promising candidates with highest Young's moduli. With the datasets curated using high-throughput CG MD simulations, we experimentally validate the simulation results with experimental mechanical characterization using AFM and scanning electron microscopy (SEM) in order to assess the viability of both the CGMD and ML approaches to modulus prediction. 
\subsection*{Mechanical properties of di- and tri-peptides from CG MD simulations}
\begin{figure}
  \includegraphics[width=\linewidth]{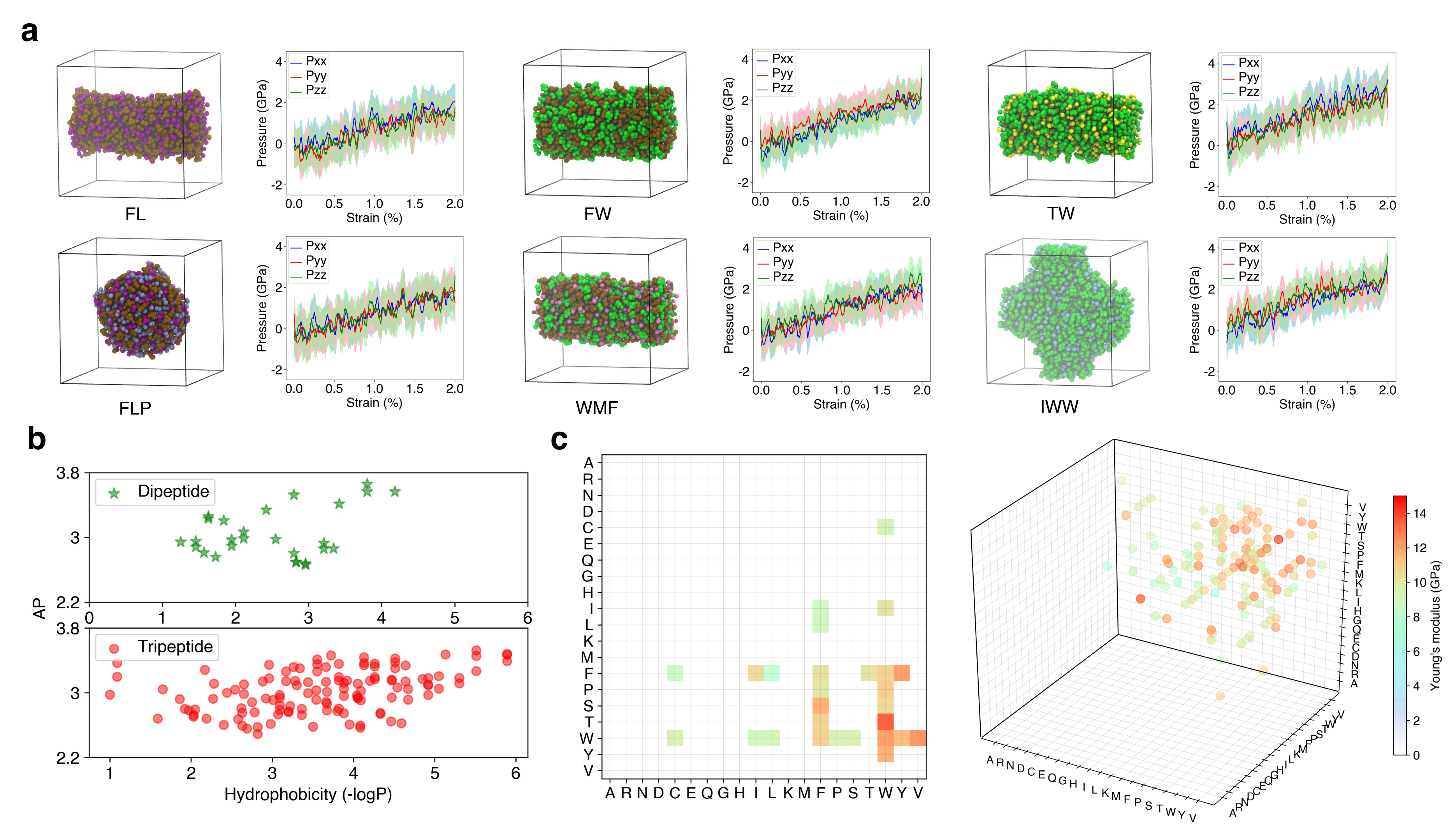}
  \caption{\textbf{Data statistics of dipeptides and tripeptides}. \textbf{a}, Example self-assembled nanostructures for dipeptides (first row) and tripeptides (second row), and corresponding strain-stress response under hydrostatic tensile test. (P$_{xx}$, P$_{yy}$ and P$_{zz}$ are pressure along \textit{x}, \textit{y} and \textit{z} directions)\textbf{b}, Aggregation propensity and hydrophobicity of dipeptides and tripeptides. \textbf{c}, Modulus statistics of dipeptides (left) and tripeptides (right) given different amino acid sequences. The colorbar is shared between both plots.}
  \label{fig:2}
\end{figure}

Building on prior research \cite{doi:10.1021/jz2010573, RN305} that employed the same CG MD methods to explore the self-assembly process of polypeptides, we selected peptide sequences with an AP>2 as those prone to assembly. We thus end up with 27 dipeptide sequences from $20^2$ possible sequences and 124 tripeptide sequences from $20^3$ possible sequence. Typical assembled nanostructures of dipeptides and tripeptides are shown in Fig. \ref{fig:2}a, where each colour represents a different amino acid. We calculate the AP and hydrophobicity values for these self-assembling peptides (as displayed in Fig. \ref{fig:2}b). The AP values obtained in this study are notably higher than those reported in previous studies \cite{doi:10.1021/jz2010573, RN305}. This increase is attributed to the use of higher peptide concentration, which ensure that the assembled structures span across the whole simulation box for mechanical tests. The hydrophobicity is always positive, reflecting the fact that all of our peptide systems aggregate in an aqueous environment, and peptide-peptide interactions are favoured over peptide-solvent interactions. \\
\\
After relaxing the peptide systems to allow the formation of assembled fibrous structures, we perform a hydrostatic tension test. Representative strain-stress responses exhibit a relatively linear curve under the small deformation, as shown in Fig. \ref{fig:2}a. These curves, from left to right, present elevated stress levels, which lead to increased Young's moduli for the peptide assemblies. All  dipeptides and tripeptides modulus values are shown in Fig. \ref{fig:2}c given different amino acid sequences. Young's modulus values of all dipeptides and tripeptides are listed in Table \ref{tab:S_dipeptide_modulus} and Table \ref{tab:S_tripeptide_modulus}, respectively. Further insights into the relationship between peptide sequence and corresponding modulus are discussed in \hyperref[sec:trends]{Summarized findings}. These findings give insights into rules that can be used for designing peptide self-assemblies with superior mechanical performances. 

\subsection*{Experimental investigation of selected di- and tri-peptides}

In order to validate the predictions made in our model, we selected 5 dipeptides and 4 tripeptides for further analysis. The peptides were chosen to represent a range of results, in terms of aggregation propensity and modulus, a summary of reasoning is presented in \ref{tab:experiemntal_reasoning}. Overall, FF, LF, PF, LA, WW, FFF, CWF, LPF and WLL were used for testing. Although there is a high proportion of aromatic amino acids in this selection, this is representative of the dipeptides with AP>2, all of which had at least one of F or W \ref{tab:dipeptide_ap}.\\
\\
\begin{figure}
  \includegraphics[width=\linewidth]{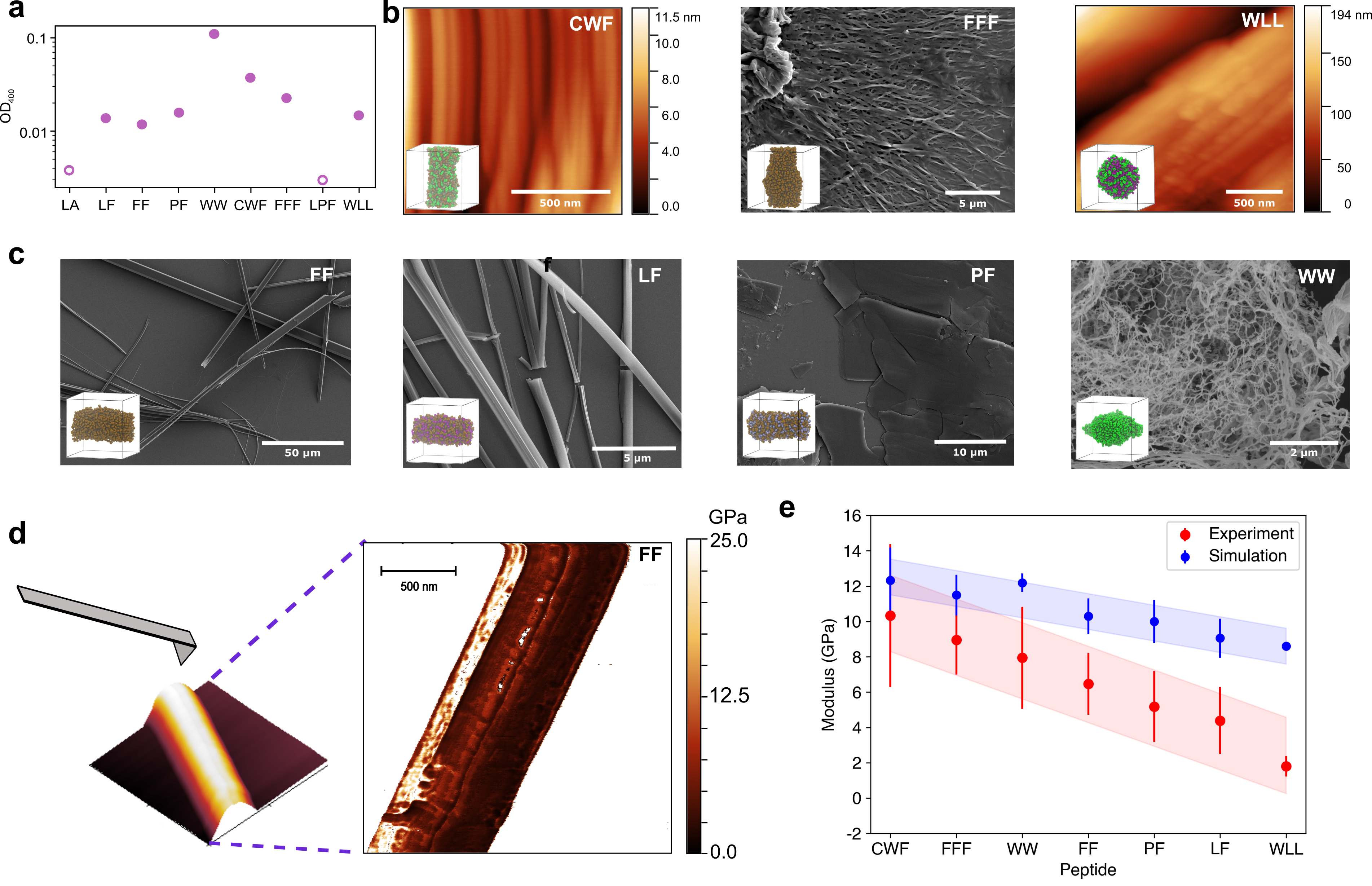}
  \caption{\textbf{Modulus validation in di- and tri-peptides. a,} Optical density measurements at 400 nm. Filled circles indicate assembly and empty circles indicate no assembly. SEM micrographs or AFM images for named, assembled tripeptides \textbf{b,} and dipeptides \textbf{c}, with corresponding CG-MD projections inset. \textbf{d}, Schematic indicating QNM-AFM methodology with a high projection of an LF fibre and corresponding modulus map. \textbf{e}, Calculated peptide modulus values from AFM experiments in comparison with simulation results.}
  \label{fig:3}
\end{figure}
To initially investigate the self-assembly behaviour of each peptide, aqueous solutions (1 mg/ml or 0.1 mg/ml for peptides insoluble at 1 mg/ml) were drop-cast and imaged using brightfield microscopy. Optical density (OD) measurements were taken (Figure \ref{fig:3}a) to quanify this behaviour. SEM micrographs were used to investigate the morphology of self-assembled peptides (Figure \ref{fig:3}b-c), and imaging confirmed that LF, FF and FFF formed fibres, PF formed sheets and WW formed a fibrous gel. CWF and WLL formed gel-like assemblies, but were deemed less stable than WW as they could not be prepared for SEM through freeze-drying as with WW. Hence, they were imaged using AFM in order to confirm a fibrous structure (Figure \ref{fig:AFM}). Brightfield imaging confirmed that LA and LPF did not self-assemble, as no clear structures were observed and these sequences had a significantly lower OD to the other peptides.\\
\\
All other short peptides (FF, PF, WW, LF, CWF, FFF) behaved as expected from our predictions, with the exception of LPF. It is interesting that both LF and PF self-assemble into fibers, while the tripeptide combining these amino acids doesn't have the same propensity to do so. This could be a reflection on the fine balance in interactions required for assembly, where the interactions can change dramatically by a simple switch of a single charge. In addition, the use of AP$>$2 as a predictor of self-assembly propensity is a binary cut-off point that may not accurately capture the more nuanced nature of self-assembly to form a defined structure, which is observed here. Interestingly, LPF has the lowest AP of all tested tripeptides (2.07), which may account for this behaviour.\\
\\
To characterise the mechanical properties of assembled peptides, and compare these to simulation values, PeakForce quantitative nanomechanical mapping (PF-QNM) measurements were performed. This technique has been widely used to determine the Young's modulus of protein amyloid fibrils from $\alpha$-synuclein $\beta$-lactoglobulin, bovine serum albumin and lysozyme.\cite{C2NR30768E,https://doi.org/10.1002/anie.201409050} Peptide assemblies were prepared as previously, and dried on a mica substrate prior to measurements. Representative height and stiffness maps are shown in Figure \ref{fig:3}d. Overall, CWF was found to have the highest modulus (10.3$\pm$8.09 GPa) of all tripeptides investigated, whereas WW was found to be the dipeptide with the highest modulus (7.94$\pm$5.78 GPa). Overall, the pattern of peptide modulus was found to be consistent with that of the prediction (Figure \ref{fig:3}e). \\
\\
Interestingly, all experimental values were consistently lower than the predicted values. There are several plausible explanations for this disparity; in the simulations the peptide concentration is significantly higher than in experiments; the simulation box is also much smaller, reducing the likelihood of structural defects being present. Further to this, there are also short-comings in both methods. Coarse grained approaches lead to loss of atomistic details, and peptide synthesis and purification can still result in some chemical impurities being present in the structure. This indicates that while using simulations to predict absolute modulus presents some challenge, the prediction of relative modulus is still valid, given the strong agreement between predicted and experimental modulus hierarchy. This finding affirms our model as a valuable approach to predicting short peptide modulus, with the potential to identify key candidates for forming stiff self-assembled structures.

\begin{figure}
  \includegraphics[width=0.8\linewidth]{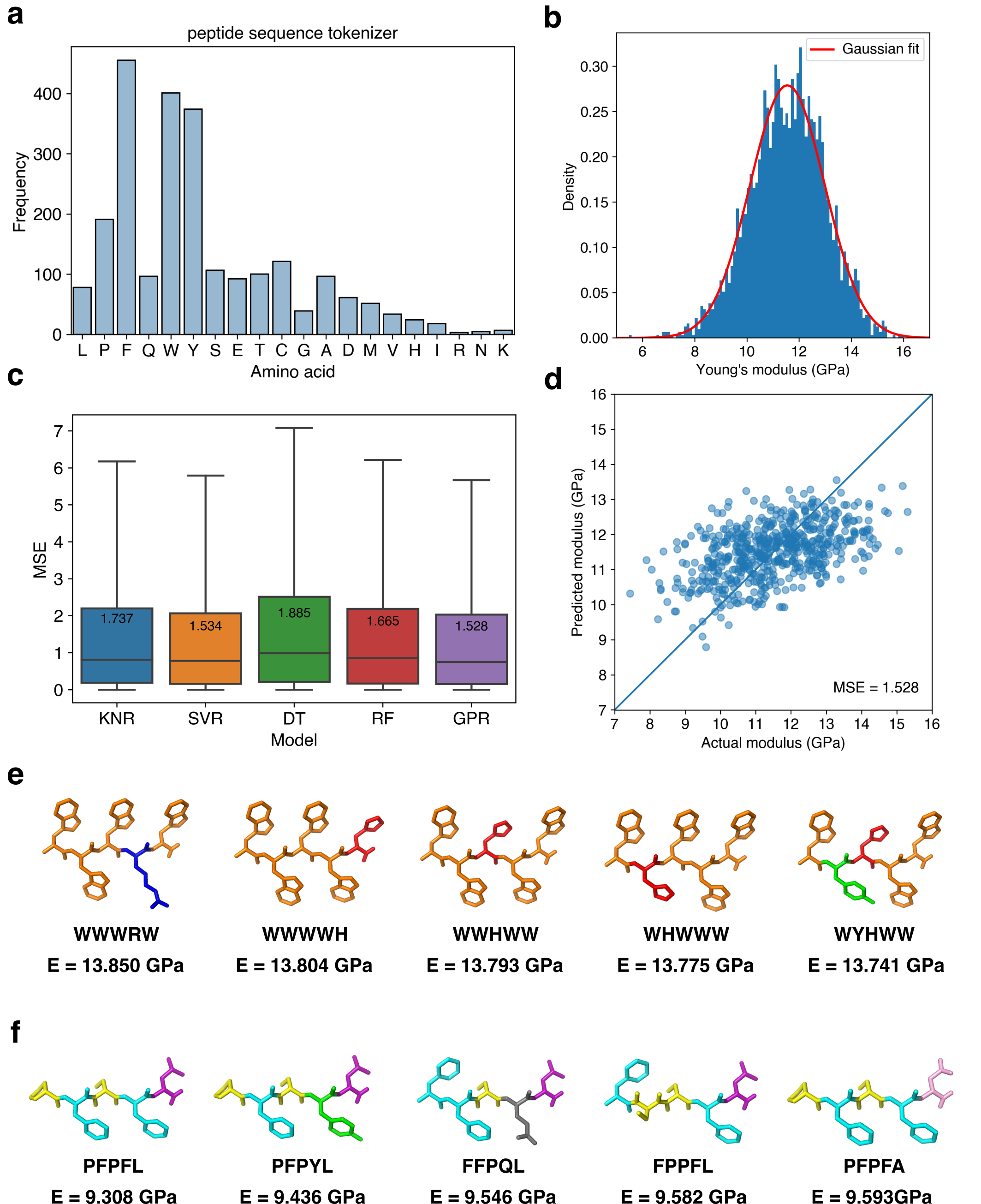}
  \centering
  \caption{\textbf{ML model for pentapeptide modulus screening}. \textbf{a}, Occurrence frequency of amino acids within the 2990 pentapeptide sequences collected. \textbf{b}, Distribution of Young's modulus for pentapeptides calculated from CG MD simulations. \textbf{c}, Performance comparison of 5 different classical ML algorithms (KNR, SVR, DT, RF, GPR) in modulus prediction given the input pentapeptide sequence. MSE is used for performance evaluation. \textbf{d}, Comparison of predicted (ML) and actual (MD) modulus with the optimal model (GPR). \textbf{e}, Top 5 pentapeptide candidates with highest Young's moduli. \textbf{f}, Bottom pentapeptide candidates with lowest Young's moduli.  }
  \label{fig:4}
\end{figure}

\subsection*{ML for pentapeptide modulus prediction}
Following validation of our simulations on di- and tripeptides, we expanded this approach to include the chemical space of pentapeptides. In this case, a full screen of the chemical space is unfeasible given that there are 3.2M possible sequences. Therefore, we selected a subset of these sequences to train a ML model that predicts the modulus for the remaining sequences. Similar to our approach with dipeptides and tripeptides, we first selected for pentapeptide sequences that can form aggregates based on prior research.\cite{Batra2022} Unlike the di- and tripeptide cases, where previous studies perform CGMD simulations to calculate the AP for all sequences, existing work on pentapeptides adopts a similar partial screening approach as this study. The APs of thousands of sequences from the literature are collected, and this data is used to train a ML algorithm for AP prediction across all pentapeptide sequences (Fig. \ref{fig:S_pentapeptide_AP}a).\cite{Batra2022} The ML model used here is a Gaussian Process Regressor (GPR), which provides a similar mean absolute error value (MAE = 0.065) as the previous study (MAE = 0.063). \cite{Batra2022} Consequently, we implemented the GPR algorithm to search for pentapeptides with AP $>$ 2. Among 3.2M possible pentapeptide sequences, the model screened for more than 25000 candidates that can self-assemble. The distribution of amino acids within these sequences is shown in Fig. \ref{fig:S_pentapeptide_AP}b.\\
\\
From this pool of self-assembling candidates, we randomly selected 2,380 pentapeptide sequences. Together with the 610 sequences previously identified \cite{Batra2022}, our dataset comprised of 2,990 unique self-assembling sequences. The occurrence frequency of the 20 amino acids in our dataset is displayed in Fig. \ref{fig:4}a. We again performed high-throughput mechanical tests to these assemblies using CG MD simulations. The Young's moduli of the pentapeptides collected showed a Gaussian-like distribution (Fig. \ref{fig:4}b). With the data, we then trained ML models to predict the Young's modulus of these pentapeptides based on their sequences. To optimize the performance of the prediction, we compared 5 different classical ML algorithms including K-nearest neighbor regressor (KNR), support vector regressor (SVR), decision tree (DT), random forest (RF) and GPR. For each algorithm, we conduct hyperparameter optimization using grid search to maximize performance. Among the 5 tested models, the GPR model outperformed any other model with lowest mean squared error (MSE) (1.528) (Fig. \ref{fig:4}c). We then further visualized the predicted modulus values against the ground truth values from CG MD simulations, revealing a high correlation (Fig. \ref{fig:4}d). The correlation observed is not as strong as that seen in AP prediction (Fig. \ref{fig:S_pentapeptide_AP}a). This discrepancy can be attributed to the self-assembling peptides being situated at the tail of the distribution from the standpoint of self-assembly behaviors, a region that is challenging for ML algorithms to accurately capture due to the highly biased and constrained input space. \\
\\
With the optimal ML model, we are able to predict the Young's modulus for over 25,000 self-assembling pentapeptides with high computational efficiency. As a result, peptide candidates with the highest and lowest modulus values can be identified from the statistics. We list top and bottom 5 pentapeptide candidates, as shown in Fig. \ref{fig:4}e and Fig. \ref{fig:4}f respectively. Similar trends to those observed in dipeptides and tripeptides are evident; for example, tryptophan is commonly found in these top pentapeptide sequences. In addition, histidine ranks as the second most frequent amino acid in predicted pentapeptides that exhibit a superior modulus. Conversely, in the bottom candidates, phenylalanine, proline and leucine emerge as the three most common amino acids, which are associated with a relatively low modulus. These observations offer more comprehensive design principles and reveal a broader range of amino acid combinations compared to shorter peptides such as dipeptides and tripeptides.  
\\

\phantomsection
\subsection*{Sequence-property relationships}\label{sec:trends}
Analysis of the generated AP and modulus values revealed valuable insights into key trends in aggregating sequences, and the relationship between specific residues and modulus values. Across all peptide lengths, hydrophobic residues, and specifically aromatic residues, have a high frequency among self-assembling peptides. However, for pentapeptides, there is a higher occurrence of non-hydrophobic residues such as serine, threonine and glutamic acid, than in di- and tri-peptides. This is consistent with the structure of peptide amphiphiles, with both hydrophobic and hydrophilic character.\cite{doi:10.1021/ja9627656} Therefore, this finding suggests that in very short sequences, hydrophobic interactions dominate, and as the chain length increases such that alternating residues can be introduced, it is the amphiphilic nature of the peptides which is a key driver for the self-assembly.\\
\\
Further examination of proline positional frequency also provides insight into the mechanisms which could drive assembly in this system. Within pentapeptides, it is significantly more likely to be found in position 3 than any other position (Fig. \ref{fig:S_proline_position}) . This suggests that the bending a proline introduces in a peptide chain promotes assembly more than having proline at any other positions. A central bend facilitates the formation of key interactions during the assembly process. The same trend is not present for the shorter peptides, and although datasets are significantly smaller for the shorter peptides, their shorter length suggests a bend is less likely to promote additional interactions in the way it does for longer peptides. This trend is consistent with findings for proteins, where the introduction of the cyclic proline facilitates protein folding. \cite{LEVITT1981251} \\
\\
Modulus values generated through the workflow show differing trends to AP data, with all 5 charged residues having normalised modulus values within the top 8 amino acids.(Fig. \ref{fig:S_modulus_frequency}) This data suggests that while charged residues aren't key to the self-assembly of these peptides, they have a strong contribution to the modulus. This finding is in contrast to the stiffest pentapeptide sequences predicted using the ML approach, which show tryptophan to be in all of the 5 stiffest pentapeptides. However, tryptophan is the fourth listed residue for contribution to a high modulus. The combination of these findings could suggest that the contribution of WW is over estimated using the ML approach, possibly due to its strong influence on assembly propensity. However, further work is required to confirm this. 

\section*{Discussion}
In this work, we systematically study the mechanical properties of peptide self-assemblies under different amino acid sequences using a combined framework of CG simulations, AFM determination and ML predictions. Given that the majority of previous studies have focused on investigating the self-assembly propensity and phases, our work addresses the gap in studying the functionalities of these assembled nanostructures. Here, we identify peptide candidates with superior stiffness across peptide lengths and summarize the key principles for designing peptide sequences. These findings offer valuable guidance for the design of peptide-based materials which can be beyond the systems studied in this work. \\
\\
The trade-off between accuracy and computational cost highlights the importance of this combined framework. From ML-based approach, to simulation, and finally to experiment, the accuracy of modulus evaluation of peptide self-assembly increases, whereas the cost and time required to perform each individual test also increases greatly. Hence, for longer sequences with a larger chemical space, ML is favoured, whereas when exhaustive screening is feasible, this is prioritised. When it comes to experimental characterization, although it is limited to testing and evaluating of only a few peptide candidates, it provides crucial validation for our high-throughput simulations. This hierarchical workflow allows us to bridge the gap between accuracy and costs. A future direction for this work could involve integrating these computational and experimental approaches into an active feedback loop. The ML models and MD simulations could serve as two filters to identify promising peptide candidates for experimental validation. Subsequently, the results from our QNM experiments could be used to enhance the datasets for training the ML model or refining the CG simulations. This iterative process will thus continuously improve prediction accuracy and efficiency.  \\
\\
Apart from the peptide amino acid sequence, a range of other factors can also play a significant role in determining self-assembly behavior and mechanical properties. These include intrinsic parameters, such as chemical modifications, \cite{Levin2014, JUNG20082143} as well as extrinsic factors such as pH, \cite{doi:10.1021/acsabm.2c00188, gels9060441} temperature \cite{doi:10.1021/ja0764862} and solvent type.\cite{doi:10.1021/jacs.0c03425, doi:10.1021/la800942n, doi:10.1021/nn404237f} A CG force field such as MARTINI can also supports simulations involving these intrinsic or extrinsic factors. Thus, by simulating a wide range of solvents including water, alcohols, benzene, and other organic solvents.\cite{martini} Therefore, a potential avenue for future research is to explore the impact of various experimental conditions on the self-assembly phases and functionalities of peptide systems through high-throughput CG MD simulations similar as the reported framework. Beyond mechanical properties, other functionalities such as photoactivity, electrical activity, and thermal properties could also be explored with the appropriate choice of force field \cite{https://doi.org/10.1002/wcms.1121,https://doi.org/10.1002/jcc.21367} or other computational methods such as Density Functional Theory (DFT).\cite{C9CS00085B} We envision that these studies will unlock numerous interesting approaches through enabling informed sequence selection when tuning the properties of peptide assemblies, while also offering comprehensive insights for related experiments.

\section*{Methods}

\phantomsection
\subsection*{CG MD simulations}\label{sec:CGMD}
To generate atomic coordinates for each peptide (di-, tri- or penta-), a template structure with repeated Glycine amino acids is first created in Pymol.\cite{PyMOL} The residues are then grafted to the template using Visual Molecular Dynamics (VMD) scripting tools \cite{HUMPHREY199633} and converted to CG representation in the MARTINI force field (version 2.2) using the open-source script martinize.py.\cite{martinize} Same as the previous studies, \cite{doi:10.1021/jz2010573, Frederix2015, Batra2022} the input flag for secondary structures is set to "E" (extended $\beta$ secondary structure) for all amino acids. \\
\\
The simulations are all executed using GROMACS codes.\cite{ABRAHAM201519} 450 (300) (180) monomers are randomly placed in a 10 \AA\ $\times$ 10 \AA\  $\times$ 10 \AA\  periodic simulation box for dipeptides (tripeptides) (pentapeptides) with standard CG water molecules. The choice of number of molecules is to keep the number of amino acids similar for peptides with different lengths. This results in a peptide concentration of 0.748 mol $L^{-1}$ (0.498 mol $L^{-1}$) (0.299 mol $L^{-1}$) for dipeptides (tripeptides) (pentapeptides). For all steps of the simulations, the distance for the Coulomb cutoff is set to 1.1 $nm$ with reaction field electrostatics. A relative dielectric constant ($\epsilon_r$) of 15 was used in standard CG water simulations for screening of the electrostatic interaction. The cutoff distance of van der Waals interaction is same as the Coulomb cutoff and the potential is shifted by a constant such that it is zero at the cutoff. Bond lengths are constrained using the LINCS algorithm.\cite{doi:10.1021/ct700200b} \\
\\
The whole CG simulation involves mainly three steps: 1) The polypeptide system is first energy-minimized for 50000 steps (time step $dt$ = 0.025 ps) until the maximum force is lower than 20 kJ $mol^{-1}$ $nm^{-1}$. 2) The simulation box is then equilibrated for $1 \times 10^7$ steps of 0.025ps using the velocity rescaling (temperature thermostat) \cite{10.1063/1.2408420} and Berendsen algorithms (pressure barostat) \cite{10.1063/1.448118} to set temperature ($\tau_T$ = 1 ps) and pressure ($\tau_P$ = 3 ps) around 303 K and 1 bar, respectively. 3) Hydrostatic tensile test is then performed to the system by expanding the box along all \textit{x}, \textit{y}, \textit{z} directions at constant speed (1 \AA/fs). The temperature is set to 303K with velocity rescaling algorithm and no pressure coupling is used during the equilibration process. The deformation step lasts for $4 \times 10^6$ steps of 0.5 fs, which results in around 2\% tensile strain at the end of simulation. The hydrostatic pressure is determined by averaging the tensile stress components ($P_{xx}$, $P_{yy}$, $P_{zz}$) across three directions, derived from executing "gmx energy" scripts. To ascertain the final mean pressure post-deformation, we uniformly sample 20 frames from the last $1 \times 10^5$ simulation steps (2.5\% of the whole deformation step). \\
\\
The GROMACS sasa tool is utilized to compute the AP values. The AP value is defined as the ratio of the solvent-accessible surface area of the structure at the beginning and end of the simulation. The hydrophobicity of a peptide is calculated using Wimley–White whole residue hydrophobicity scales. \cite{WHITE1998339, doi:10.1021/bi9600153} The Wimley-White whole residue hydrophobicity scales show the free energy of transfer from water to n-octanol. The formula is as follows:
\begin{flalign}
&& \log P = \sum_{r} \Delta G_{\text{woct}, r} && 
\label{eq:logP}
\end{flalign}
where $r$ stands for "residue". $\Delta G_{woct, r}$ is the free energy difference from water to n-octanol.\\
\\
To calculate the volume of assembled nanostructures, we perform a Monte Carlo-based random sampling approach. This approach involves distributing random "seeds" throughout the simulation box. Each seed is then evaluated to determine whether it lies within the boundary of the nanostructure by checking the proximity of atoms around the seed given a cutoff distance of 5 \AA\  . If a seed is surrounded by atoms, it is counted as part of the structure. By calculating the proportion of seeds that fall within the assembled nanostructure to the total number of seeds distributed, we can estimate the volume occupied by the the fibrous structure. Given the cylindrical structure of peptide self-assemblies, the selection of cutoff distance does not affect the results as we obtain the final radius of the structure by subtracting the cutoff distance from the radius calculated through the random sampling process.

\phantomsection
\subsection*{Modulus calculation of polypeptide self-assemblies}\label{sec:modulus_polypeptide}
To calculate the Young's modulus of fibrous structures of polypeptide self-assemblies, we base ourselves on the energy conservation law. In hydrostatic tensile test, when the simulation box expands, there are two factors contributing to the total work: one is from the stretching of assembled nanostructures; the other is due to expansion of water box. The fibrous nanostructures mainly expand along the axial direction as the relative volume change is close to the tensile strain value or the stretch ratio ($lambda$ = 0.02, see Fig. \ref{sec:modulus_polypeptide}). As a consequence, the total work consists of:
\begin{flalign}
    && P_{\text{tot}}dV_{\text{tot}} = P_{\text{w}}dV_{\text{w}} + P_{\text{s}}dV_{\text{s}} \Rightarrow P_{\text{tot}}dV_{\text{tot}} = P_{\text{w}}dV_{\text{w}} + E_{\text{s}}V_{\text{s}} \lambda^2 &&
\label{eq:energy_polypeptide}
\end{flalign}

\noindent where $P$ is the pressure, $V$ is the volume, $N$ is the molecule number, $E$ is the Young's modulus, $\lambda$ is the elongation (= 2\%), the subscript “tot” stands for ”total”, “w” stands for water, “s” stands for self-assembly. The volume of assembled nanostructure is obtained by excluding the volume of water molecules from the total volume of a simulation box which is discussed below.  \\
\\
The Equation \ref{eq:energy_polypeptide} above can be further simplified given that the deformation is small. Therefore, we have the following approximation:
\begin{flalign}
    && dV_{\text{tot}} = V_{\text{tot}}((1 + \lambda)^3 - 1) \approx 3\lambda V_{\text{tot}} &&
\end{flalign}

\noindent To evaluate the influence of water molecules, we run a control simulation with a box filled only with equilibrated water molecules. The box is deformed in the same way as the box with polypeptides. When we calculate the contribution of water molecules, we utilize the pressure obtained from the control simulation after deformation ($P_{\text{w}}^0$). In terms of the occupied volume of water molecules, it is calculated based on water density from control simulation. More specifically, the volume ratio of water molecule ($r_w$) is:  
\begin{flalign}
    && r_{\text{w}} = \frac{N_{\text{w}} V_{\text{w}}^0}{N_{\text{w}}^0 V_{\text{tot}}} &&
\end{flalign}
\begin{flalign}
    && dV_{\text{w}} = 3\lambda V_{\text{tot}}r_{\text{w}} &&
\label{eq:modulus_polypeptide}
\end{flalign}

\noindent where superscript “0” refers to the control equilibrated water box. We can then plug in these individual terms to Equation \ref{eq:energy_polypeptide} to obtain the following formula for the Young's modulus:
\begin{flalign}
  && \Rightarrow E_{\text{s}} = \frac{3(P_{\text{tot}} - P_{\text{w}}^0r_{\text{w}}) V_{\text{tot}}}{\lambda^2 V_{\text{s}}} = \frac{3(P_{\text{tot}} - P_{\text{w}}^0r_{\text{w}})}{\lambda (1-r_{\text{w}})} &&
\end{flalign}

\noindent For those non-fiber assemblies like spheres, this derivation should actually be performed using the bulk modulus instead of Young's modulus. The second term in Equation \ref{eq:energy_polypeptide} should be corrected as $K_{\text{s}}V_{\text{s}} (3\lambda)^2$ However, given these spherical structures possess lower moduli and are generally not systems of interest for mechanical applications as they do not form networks with mechanical support, we here utilize the same formula (Equation \ref{eq:modulus_polypeptide}) for all structures. 

\phantomsection
\subsection*{ML models for AP and modulus prediction}\label{sec:ml_pentapeptide}
To predict AP and modulus for pentapeptides from their amino acid sequences, we train and compare a few classical ML algorithms including K-nearest neighbor regressor (KNR), support vector regressor (SVR), decision tree (DT), random forest (RF) and Gaussian process regressor (GPR). In the pentapeptide sequence, each amino acid is converted into a one-hot encoding, which represents each type of amino acid as a unique orthogonal numerical vector. We then merge these one-hot encodings for the five amino acids in the pentapeptide into a single concatenated vector used as the input for ML models. Given that there are 20 distinct natural amino acids, the input feature dimension for pentapeptides is 100. The output is either AP or modulus value. \\
\\
To optimize the performance of 5 ML models, we vary the combinations of their hyperparameters and obtain the optimal prediction accuracy for each kind of model based on the MSE metrics, a process known as the grid search. The selection of hyperparameters and the optimal combinations for the 5 tested model are listed in Table \ref{tab:S_hyperparameter}. Cross validation is implemented to evaluate models' performances under different combinations of hyperparameters using the scikit-learn package.\cite{scikit-learn} The technique is a statistical method that divides the data into several subsets, training the model on some subsets and validating it on others, to ensure it generalizes well to new data.  \\
\\
Among the 5 optimized models, we discover that the GPR model outperforms the other four algorithms by achieving the lowest MSE values. Therefore, we utilize this model for predicting both AP and Young's modulus of pentapeptides. In the AP prediction, we utilize more than 6,600 data points of random pentapeptide sequences from the earlier study \cite{Batra2022}, to train, validate and test the GPR model. For modulus prediction, our CG MD data on 2,990 self-assembling peptides form the training, validation and testing sets. In both cases, the data is divided into training, validation, and testing sets with split ratios of 60\%, 20\%, and 20\%, respectively. All the training and inference are performed on a single NVIDIA Quadro RTX 5000 GPU core (16GB) using the scikit-learn package \cite{scikit-learn} and TensorFlow library.\cite{tensorflow2015-whitepaper}. 

\phantomsection
\subsection*{Peptide synthesis}
Diphenylalanine (Bachem, Switzerland), proline-phenylalanine (Bachem, Switzerland), leucyl-phenylalanine (MedChemExpress, USA) and leucyl-alanine hydrate (Merck, Germany) were purchased from named suppliers. All other peptides were synthesised using solid state peptide synthesis and standard 9-fluorenylmethoxycarbonyl (Fmoc) chemistry on an automated microwave peptide synthesiser (Liberty Blue, CEM). Peptide were synthesised on a Wang resin with a hexafluorophosphate benzotriazole tetramethyl uronium (HBTU) (source) activator and a N,N-Diisopropylethylamine (DIPEA) activator base. After synthesis, the peptide was cleaved from the resin using buffer (2x25 mL, 2x1 hr shaking) (95\% trifluoroacetic acid (TFA), 2.5 \% triisopropyl silane (TIPS) and 2.5\% water). If tryptophan (W) was present, a buffer of 95\% TFA, 3\%TIPS, 2\% 2,2-(Ethylenedioxy)diethanethiol (DODT) and 1\% water was used to prevent oxidation. The mixture was filtered and the cleavage mixture evaporated off. The resulting viscous yellow liquid was washed with 25 mL of dichloromethane, the 3x50mL of cold diethyl ether. Liquid-Chromatography Mass Spectroscopy (LCMS) for each peptide is presented in figures \ref{fig:MS_CWF} \ref{fig:MS_FFF}, \ref{fig:MS_LPF} and \ref{fig:MS_WW}. 

\phantomsection
\subsection*{Optical density}
Optical density measurements quantifies whether a peptide is aggregated. These measurements were taken using absorbance at 600 nm with a CLARIOstar microplate reader (BMG Labtech).

\phantomsection
\subsection*{Imaging}
Scanning electron microscopy is used to visualize the peptide assemblies with sub-micron resolution. The crude peptide was dissolved in water (0.1 mg/ml), sonicated and dropcast on a glass surface. Samples were initially imaged using a brightfield microscope. Scanning electron microscope (SEM) micrographs were obtained using either a MIRA3 FEG-SEM or a CLARA2 FEG-SEM (TESCAN, Brno, Czech Republic). Samples were drop-cast and dried on a silicon wafer before being coated with 10nm Pt prior to imaging using a Q150T ES coater (Quorum Technologies, Lewes, UK). In cases where the sample showed gel-like behaviour, the peptide solutions were plunge-frozen and freeze-dried to retain the fibrillar structure and remove liquid before being coated and imaged. This method minimises collapse of the porous structure when it is exposed to vacuum for imaging. Images were taken with an SE detector (Everhart-Thornley Type) at 5 keV. Images were analysed using ImageJ.

\phantomsection
\subsection*{Nano-Mechanical AFM}
To find the modulus of the peptide assemblies experimentally, PeakForce quantative nanomechanical mapping performed on a MultiMode 8 (Bruker, USA) atomic force microscope.  Peptide solutions (0.1 mg/ml and 1 mg/ml) were dropcast onto clean Mica slides and dried at room temperature. Bruker 0.01-0.025 Ohm-cm Antimony (n) doped Si probes were used, with a spring constant of 40 N/m, a cantilever resonance frequency of 300 kHz and a nominal tip radius of 8nm. These tips were selected because their spring constant is within a comparable range to the predicted modulus values for our peptide assemblies. \\
\\
The samples were then imaged with a PeakForce setpoint of 1kHz, and the feedback gain, tip velocity, peak force amplitude and tip height were tuned to optimize sample tracking. A thermal tune was used to find the exact spring constant of the probe. Measurements were taken on 3-4 selected regions for each peptide at room temperature to obtain the topography information on the sample as well as deformation, peak force error and stiffness maps. Stiffness maps representing the Young's modulus were generated by the software by fitting the unloading curve using the DMT (Derjaguin–Muller–Toporov) model. This model describes the adhesion between 2 hard elastic spheres (the tip and the surface) within attractive distances. The reduced Young's modulus of the sample can be found by fitting the retraction curve to the DMT model, given by:
\begin{flalign}
&& E^{*} = \frac{3 \left( F - F_{\text{adh}} \right)}{4 \sqrt{Rd^{3}}} &&
\label{eq:logP}
\end{flalign}

where E\textsuperscript{*} is the reduced Young's Modulus, F is the force on the tip, F\textsc{adh} is the adhesion force, R is the tip radius and d is the tip-sample separation.\cite{DERJAGUIN1975314} Stiffness maps generated with this model from retraction curves were visualized and analyzed using Gywddian. Height maps were masked to extract the highest areas of the image, which represent the sample, then transferred to stiffness maps, and statistical analysis functions were perform to find the average modulus and standard deviation for each masked area. Areas were also selected by comparison with deformation data, in order to accurately obtain the modulus of peptide assemblies and exclude data representing the Mica surface below. The average RMSD was also calculated using Gwyddian and is presented as error bars.

\bibliography{references}

\section*{Acknowledgements}

The authors acknowledge the support from the following funding: US Department of Agriculture 2021-69012-35978 (M.J.B.); Department of Energy-SERDP WP22-S1-3475 (M.J.B.); Army Research Office 79058LSCSB, W911NF-22-2-0213, and W911NF2120130 (Z.Y. and M.J.B.); National Institutes of Health U01EB014976 and R01AR077793 (M.J.B.); Office of Naval Research N00014-19-1-2375 and N00014-20-1-2189 (M.J.B.); Engineering and Physical Sciences Research Council Cambridge Nanoscience and Nanotechnology EP/S022953/1 (S.K.Y., J.O.B., A.R.); Cambridge Display Technology Ltd. (S.K.Y.); European Research Council under the European Union’s Seventh Horizon 2020 research and innovation program through the ERC grant DiProPhys 101001615 (T.P.J.K.); and Biotechnology and Biological Sciences Research Council (T.P.J.K.).

\section*{Author contributions statement}

M.J.B., T.P.J.K., S.K.Y. and Z.Y. conceived the idea. Z.Y. performed the CG MD simulation and developed the ML model. S.K.Y. carried out synthesis and AFM experiments. S.K.Y, A.R. and J.O.B. performed imaging. M.J.B. and T.P.J.K. supervised the project, analyzed the results, and interpreted them with S.K.Y., Z.Y. and A.L. S.K.Y, Z.Y., A.L., M.J.B. and T.P.J.K. wrote the manuscript.

\textbf{Accession codes}: Dataset and codes created and used in this study can be found at: \href{https://github.com/lamm-mit/}{https://github.com/lamm-mit/PeptideModulus}; \textbf{Competing interests}: The authors declare that they have no competing interests.  

\newpage
\begin{center}
 \LARGE\bfseries Supporting Information: High-throughput Screening of the Mechanical Properties of Peptide Assemblies   
\end{center}

%
%
%
%
\renewcommand{\thefigure}{S\arabic{figure}}
\renewcommand{\thetable}{S\arabic{table}}

\begin{figure}[htbp]
  \includegraphics[width=\linewidth]{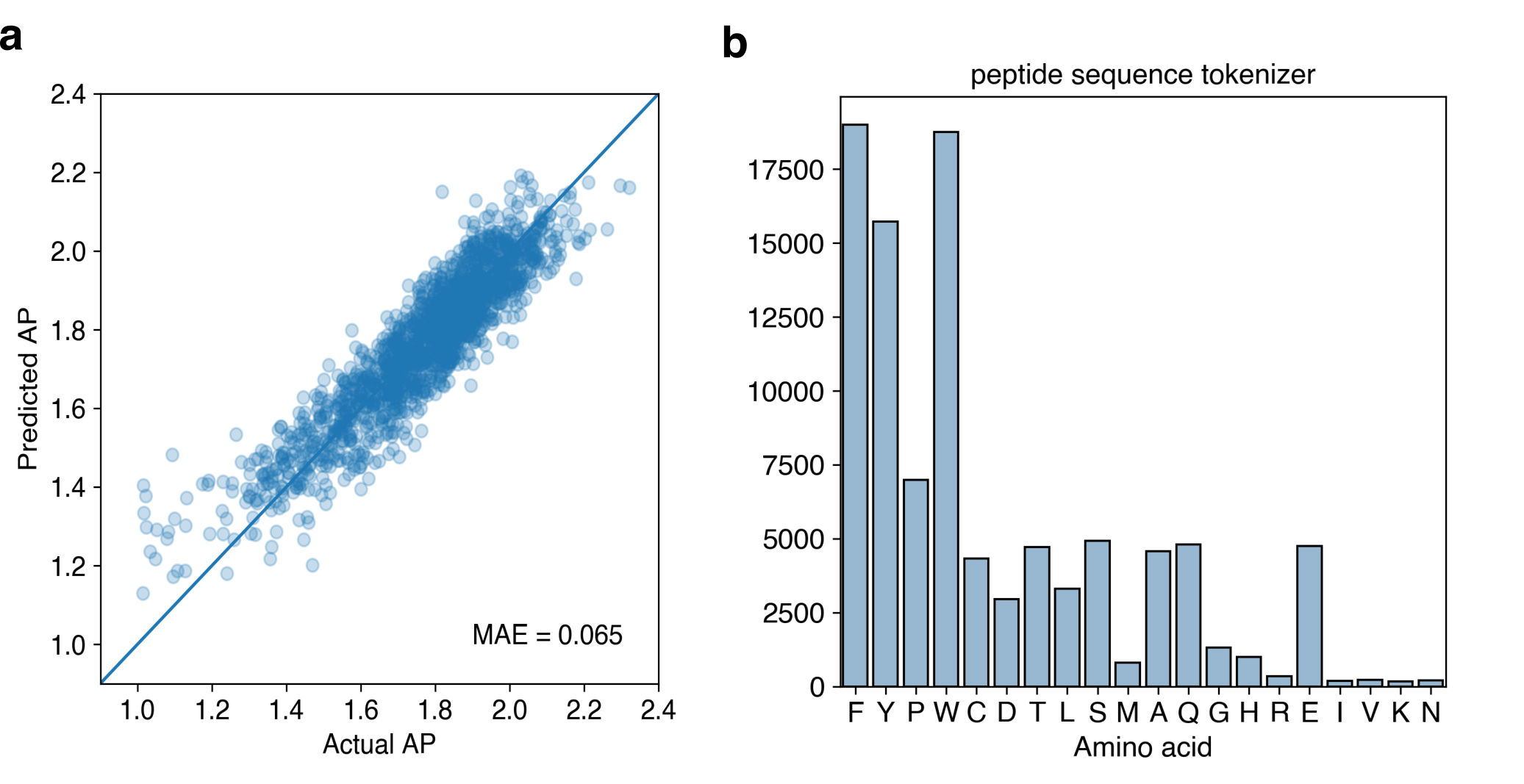}
  \caption{\textbf{Prediction of AP for pentapeptides}. \textbf{a}, Comparison of predicted and actual AP values using GPR model \textbf{b}, Occurrence frequency of amino acids within the 25328 pentapeptide sequences with AP $>$ 2.}
  \label{fig:S_pentapeptide_AP}
\end{figure}
\begin{figure}[htbp]
  \includegraphics[width=\linewidth]{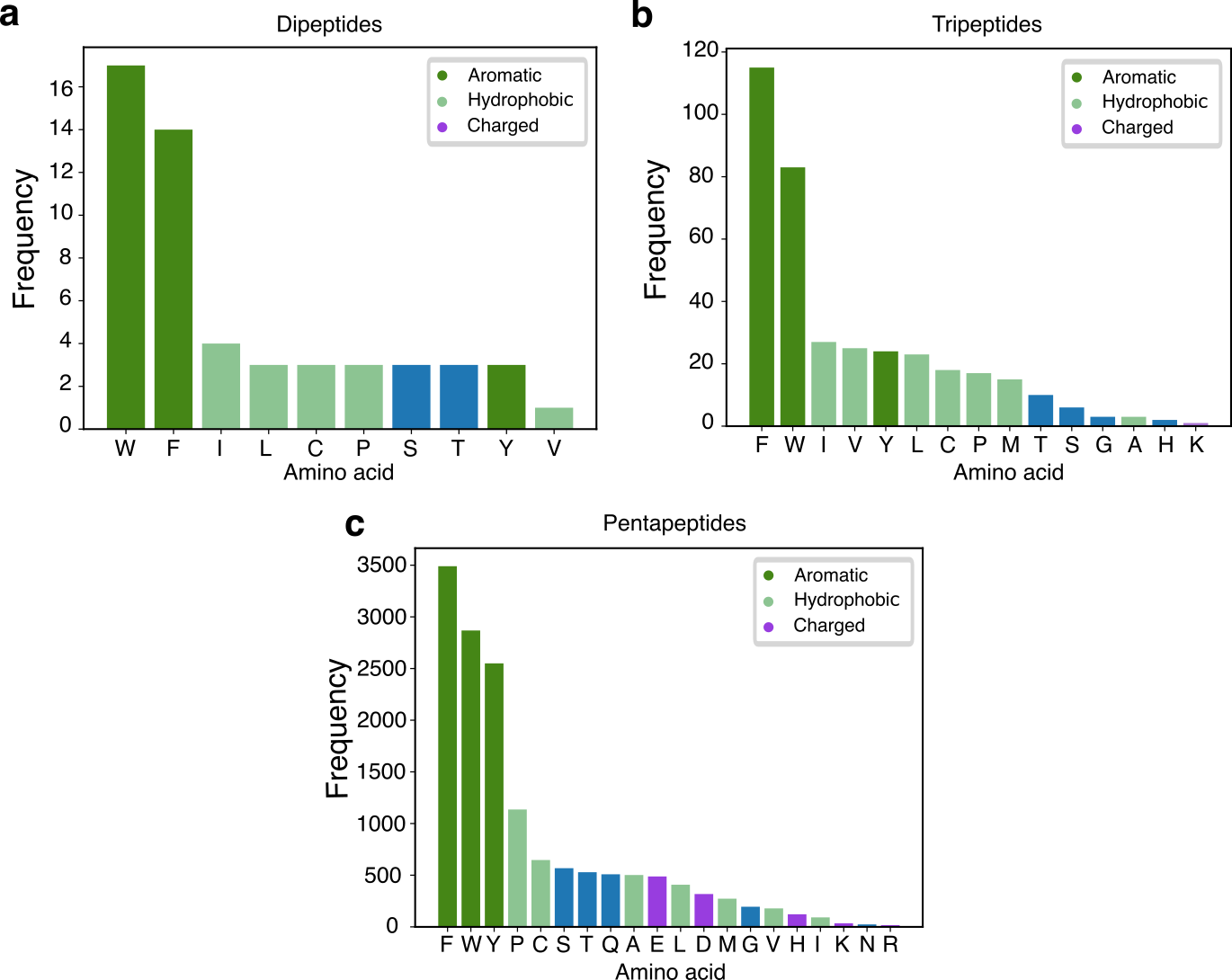}
  \caption{\textbf{Frequency of each amino acid in aggregating peptides}. \textbf{a}, dipeptides \textbf{b}, tripeptides and \textbf{c}, pentapeptides.}
  \label{fig:S_aa_frequency}
\end{figure}
\begin{figure}[htbp]
  \centering
  \includegraphics[width=0.8\linewidth]{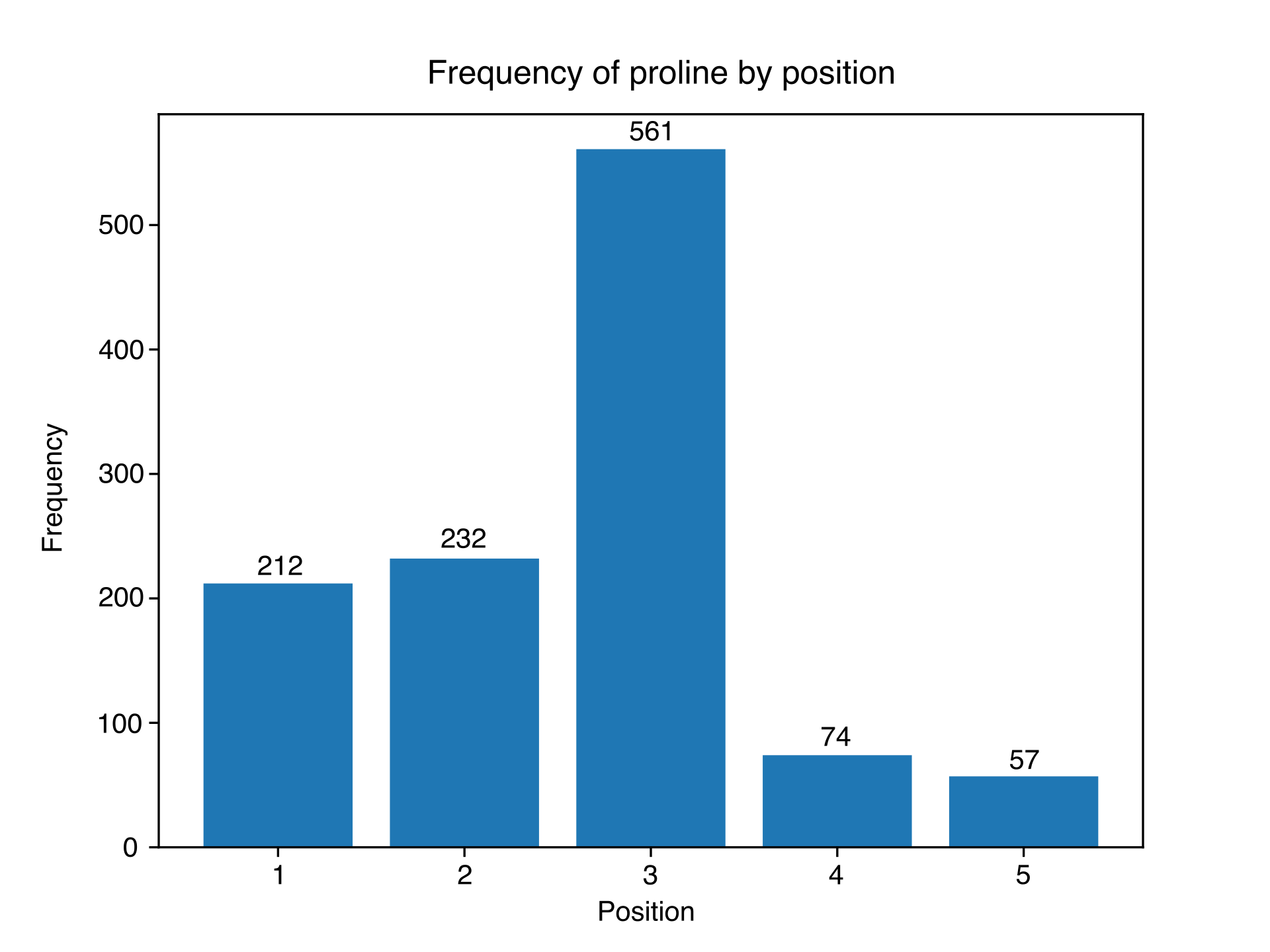}
  \caption{\textbf{Frequency of proline in each position of a pentapepeptide}.}
  \label{fig:S_proline_position}
\end{figure}

\clearpage
\begin{table}[htbp]
\centering
\caption{\textbf{Aggregation Propensity of dipeptides.} The table shows the list of aggregation rpopensity of 27 dipeptides calculated from CG MD simulations. Reproduced from \cite{doi:10.1021/jz2010573}. AP for tripeptides can be found here \cite{Frederix2015}.\label{tab:dipeptide_ap}}

\resizebox{\textwidth}{!}{%
\begin{tabular}{cc|cc|cc}
\hline
\textbf{Dipeptide} & \textbf{Aggregation propensity} & \textbf{Dipeptide} & \textbf{Aggregation propensity} & \textbf{Dipeptide} & \textbf{Aggregation propensity} \\ \hline
FL                 & 2.0              & LF                 & 2.4             & PW                 & 2.0           \\
FC                 & 2.0              & FT                 & 2.3             & FF                 & 3.2             \\
IF                 & 2.3              & IW                 & 2.2             & WY                 & 2.1             \\
WI                 & 2.0              & FI                 & 2.0             & YW                 & 2.2             \\
WL                 & 2.0              & SW                 & 2.6             & SF                 & 2.3              \\
CW                 & 2.4              & FW                 & 3.5             & FY                 & 2.2             \\
WP                 & 2.0              & PF                 & 2.1             & WV                 & 2.1             \\
WS                 & 2.4              & WF                 & 3.5             & WW                 & 3.2             \\
WC                 & 2.4              & TF                 & 2.1             & TW                 & 2.4             \\ \hline
\end{tabular}%
}
\end{table}
\clearpage

\renewcommand{\thetable}{S\arabic{table}}
\renewcommand{\arraystretch}{1.25} 
\clearpage
\begin{table}[htbp]
\centering
\caption{\textbf{Young's modulus of dipeptides.} The table shows the list of Young' moduli of 27 dipeptides calculated from CG MD simulations. \label{tab:S_dipeptide_modulus}}

\resizebox{\textwidth}{!}{%
\begin{tabular}{cc|cc|cc}
\hline
\textbf{Dipeptide} & \textbf{Young's modulus (GPa)} & \textbf{Dipeptide} & \textbf{Young's modulus (GPa)} & \textbf{Dipeptide} & \textbf{Young's modulus (GPa)} \\ \hline
FL & 7.807720540949371 & WI & 9.012563928086287 & WP & 9.393206496984947 \\ 
FC & 8.722461101186608 & WL & 9.174486069657444 & WS & 9.546606404885232 \\ 
IF & 8.840932968157214 & CW & 9.269547925071892 & WC & 9.554193046269539 \\ 
LF & 9.052405708879773 & PF & 9.998119628019074 & FT & 10.169981742601308 \\ 
IW & 10.235060944275526 & FF & 10.29099566718512 & FI & 10.458761516385202 \\ 
SW & 10.547544995480502 & FW & 10.799657966229544 & WF & 10.868993774120387 \\ 
TF & 10.872259750331786 & PW & 10.974864846174846 & WY & 11.500974591520219 \\ 
YW & 11.743708009190165 & SF & 11.82538728553692 & WW & 12.193461950752054 \\ 
FY & 12.205059345714034 & WV & 12.451316519260242 & TW & 13.503187828127807 \\ \hline
\end{tabular}%
}
\end{table}
\clearpage
\begin{table}[htbp]
\centering
\caption{\textbf{Young's modulus of tripeptides.} The table shows the list of Young's moduli of 124 tripeptides calculated from CG MD simulations.\label{tab:S_tripeptide_modulus}}

\resizebox{\textwidth}{!}{%
\begin{tabular}{cc|cc|cc|cc}
\hline
\textbf{Tripeptide} & \textbf{\begin{tabular}[c]{@{}c@{}}Young's modulus \\ (GPa)\end{tabular}} & \textbf{Tripeptide} & \textbf{\begin{tabular}[c]{@{}c@{}}Young's modulus \\ (GPa)\end{tabular}} & \textbf{Tripeptide} & \textbf{\begin{tabular}[c]{@{}c@{}}Young's modulus \\ (GPa)\end{tabular}} & \textbf{Tripeptide} & \textbf{\begin{tabular}[c]{@{}c@{}}Young's modulus \\ (GPa)\end{tabular}} \\ \hline
LPF & 6.340702467126743 & GFF & 9.468811163142984 & PWI & 10.380258122105067 & WVF & 11.639484435902789 \\
LCF & 6.919641188564055 & GFW & 9.487866870693468 & WCW & 10.398082843145888 & CFF & 11.672581544309521\\
CPW & 7.22065225980759 & YFF & 9.555165191202033 & TYF & 10.423527446085737 & PWF & 11.766233943284302 \\
MIF & 7.414686607156735 & LWY & 9.572585247550304 & FWF & 10.531507816287872 & VYF & 11.82208873527942 \\
IIF & 7.951676203850698 & LYF & 9.617955133716093 & MFF & 10.617261004798305 & FCW & 11.860555833010899 \\
IMW & 8.059834868914397 & IFF & 9.628398231911584 & LFF & 10.620969160462 & FFT & 11.86126185905201 \\
WFG & 8.089308415961293 & WFL & 9.634018119783613 & FFM & 10.694814105957446 & WFF & 11.879041683530355 \\
IIW & 8.290137158819599 & PMW & 9.635455051260976 & YWF & 10.715807956834436 & TWF & 12.033209072713053 \\
ILF & 8.307795334182755 & TCW & 9.654180574405153 & IYY & 10.769457267203626 & FYI & 12.03597787126243 \\
FFC & 8.314979008585954 & PLW & 9.669102432559413 & LWV & 10.836280449528331 & YFW & 12.117499544983481 \\
LWI & 8.465353896977845 & CWM & 9.686646733583403 & FYL & 10.896518686407306 & VWW & 12.167348802187256 \\
FLP & 8.47315494100645 & VPF & 9.722423726607984 & FFY & 10.910007091017503 & CWF & 12.193461950752054 \\
IFP & 8.48075999927675 & VAW & 9.751257030045872 & FWV & 10.956590088996252 & FHF & 12.270736574942424 \\
PVF & 8.515188034137749 & PFY & 9.805204187352281 & FFW & 10.985161873408591 & VYW & 12.291343177956799 \\
WLL & 8.59899909477708 & IWI & 9.81942509091995 & VFW & 10.990198488263966 & SFF & 12.31248374943476 \\
TFV & 8.76094747847812 & FWI & 9.852920221252651 & FWP & 10.991828906252447 & WFP & 12.31697592588758 \\
WFI & 8.808762777798119 & VFF & 9.905251294159283 & FWC & 10.997436377448828 & KWF & 12.319692270900246 \\ 
MFY & 8.874720923170353 & IFY & 9.927299203376228 & FFV & 11.019976680454828 & FYF & 12.322925817008102 \\
CLW & 9.004008118984038 & SCW & 9.971606797477305 & FFA & 11.071803136149539 & IYW & 12.32852945173539 \\
FFS & 9.07978201644674 & MFW & 10.002739543552156 & WLC & 11.138000125800696 & WFT & 12.328852271416231 \\
IVW & 9.14094289820765 & FVF & 10.01483291697527 & IFW & 11.180098754946592 & IFT & 12.489077603271737 \\
VVF & 9.183598291186645 & LWF & 10.119532189836342 & FWY & 11.240029820538698 & WLT & 12.53362673638183 \\
CMF & 9.270596491891055 & YFC & 10.144903833049145 & YFM & 11.259359950959645 & MWW & 12.60200015105512 \\
FLF & 9.313944401758295 & HFF & 10.151597863169224 & TFW & 11.341150690861069 & WWF & 12.794834261605867 \\
PWL & 9.328473398190361 & MYF & 10.165891393089531 & WPW & 11.404067806109047 & TFF & 12.796683964189185 \\
PCF & 9.332041731950875 & YLW & 10.177786058726161 & CFV & 11.432994633385217 & SFW & 12.913328313051876 \\
WIW & 9.342747187017478 & WVW & 10.230411106358048 & IWF & 11.433904581999618 & FWW & 12.961673954958233 \\
LIF & 9.348157034421874 & ISW & 10.271148719373835 & VMW & 11.461230052977898 & AFW & 12.993185573455081 \\
LYW & 9.392357318468154 & PFF & 10.301337046688314 & FFF & 11.48955075156933 & VWF & 13.362853318282255 \\
WCF & 9.40146062820604 & VWV & 10.374953164108293 & WFW & 11.48986373487116 & IWW & 13.366584119713385 \\
WFM & 9.439357913666969 & WMF & 10.378637050745665 & FVV & 11.514835399113373 & SWW & 14.0178122087803 \\ \hline
\hline
\end{tabular}%
}
\end{table}
\clearpage
\begin{table}[htbp]
\centering
\caption{\textbf{Hyperparameter selection}. In the table, we list the selected hyperparameters of each type of model used in the grid search cross validation for hyperparameter optimization. The grid search of hyperparameters is done using the scikit-learn package\cite{scikit-learn} and the choices of hyperparameters follow the syntax of the package. The optimal values of various hyperparameters for different ML models are highlighted in bold.\label{tab:S_hyperparameter}}

\resizebox{\textwidth}{!}{%
\begin{tabular}{cccccccccc}
\hline
\multicolumn{2}{c}{\textbf{KNR}}                                                                                                                        & \multicolumn{2}{c}{\textbf{SVR}}                                                                                                                          & \multicolumn{2}{c}{\textbf{DT}}                                                 & \multicolumn{2}{c}{\textbf{RF}}                                                                                                   & \multicolumn{2}{c}{\textbf{GPR}}                                                                                                                                                                \\ \hline
Parameter                                                  & Value                                                                                      & Parameter                                                           & Value                                                                               & Parameter & Value                                                               & Parameter                                                   & Value                                                               & Parameter                & Value                                                                                                                                                                \\ \hline
\begin{tabular}[c]{@{}c@{}}\# of \\ neighbors\end{tabular} & \begin{tabular}[c]{@{}c@{}}1, \\ 2, \\ 5, \\ \textbf{10}\end{tabular}                               & kernel                                                              & \begin{tabular}[c]{@{}c@{}}\textbf{'linear'}, \\ 'poly', \\ 'rbf', \\ 'sigmoid'\end{tabular} & max depth & \begin{tabular}[c]{@{}c@{}}2, \\ \textbf{3}, \\ 5, \\ 10, \\ 20\end{tabular} & max depth                                                   & \begin{tabular}[c]{@{}c@{}}2, \\ 3, \\ 5, \\ \textbf{10}, \\ 20\end{tabular} &                          &                                                                                                                                                                      \\
{\color[HTML]{000000} weights}                             & \begin{tabular}[c]{@{}c@{}}'uniform', \\ \textbf{'distance'}\end{tabular}                           & degree                                                              & \begin{tabular}[c]{@{}c@{}}\textbf{3}, \\ 6, \\ 9, \\ 12\end{tabular}                        & splitter  & \begin{tabular}[c]{@{}c@{}}\textbf{'best'}, \\ 'random'\end{tabular}         & \begin{tabular}[c]{@{}c@{}}\# of \\ estimators\end{tabular} & \begin{tabular}[c]{@{}c@{}}10, \\ 100, \\ \textbf{200}\end{tabular}          & \multirow{-2}{*}{Kernel} & \multirow{-4}{*}{\begin{tabular}[c]{@{}c@{}}None, \\ DotProduct(), \\ WhiteKernel(), \\ RBF(), \\ ExpSineSquared(), \\ \textbf{DotProduct() +} \\ \textbf{WhiteKernel()}\end{tabular}} \\
algorithm                                                  & \begin{tabular}[c]{@{}c@{}}\textbf{'auto'}, \\ 'ball\_tree', \\ 'kd\_tree', \\ 'brute'\end{tabular} & \begin{tabular}[c]{@{}c@{}}regularization \\ parameter\end{tabular} & \begin{tabular}[c]{@{}c@{}}\textbf{1}, \\ 10, \\ 100, \\ 1000\end{tabular}                   & criterion & \begin{tabular}[c]{@{}c@{}}'mse', \\ \textbf{'mae'}\end{tabular}             & criterion                                                   & \begin{tabular}[c]{@{}c@{}}'mse', \\ \textbf{'mae'}\end{tabular}             &                          &                                                                                                                                                                      \\ \hline
\end{tabular}%
}
\end{table}
\clearpage
\begin{table}[ht]
\caption{\textbf{Selection reasoning for experimentally investigated peptides.}\label{tab:experiemntal_reasoning}}

\centering
\begin{tabular}{c|c|c|l}
\hline
\textbf{Sequence} & \textbf{AP} & \textbf{Predicted modulus (GPa)} & \textbf{Selection Reasoning} \\ \hline
FF                & 3.2          & 11.146464392993568               & Extensively explored and can be used as a reference to the literature \\ 
LA                & -            & -                                & As a negative control, it isn't predicted to assemble \\ 
LF                & 2.4          & 10.130119055259193               & Includes a less hydrophobic residue \\
PF                & 2.1          & 10.823882304915925               & An intermediate modulus value \\
WW                & 3.2          & 12.476938245501941               & Represents one of the highest modulus values \\
CWF               & 2.08         & 14.942268847942001               & Represents the highest modulus value of all tripeptides \\ 
FFF               & 2.26         & 10.76086530990813                & Previously explored and an interesting comparison to FF \\
LPF               & 2.07         & 6.340702467126743                & The lowest modulus values, and a low AP.\\
WLL               & 2.18         & 8.78829076532965                 & An intermediate modulus value. \\ \hline

\end{tabular}
\end{table}

\begin{figure}[htbp]
  \includegraphics[width=0.65\linewidth]{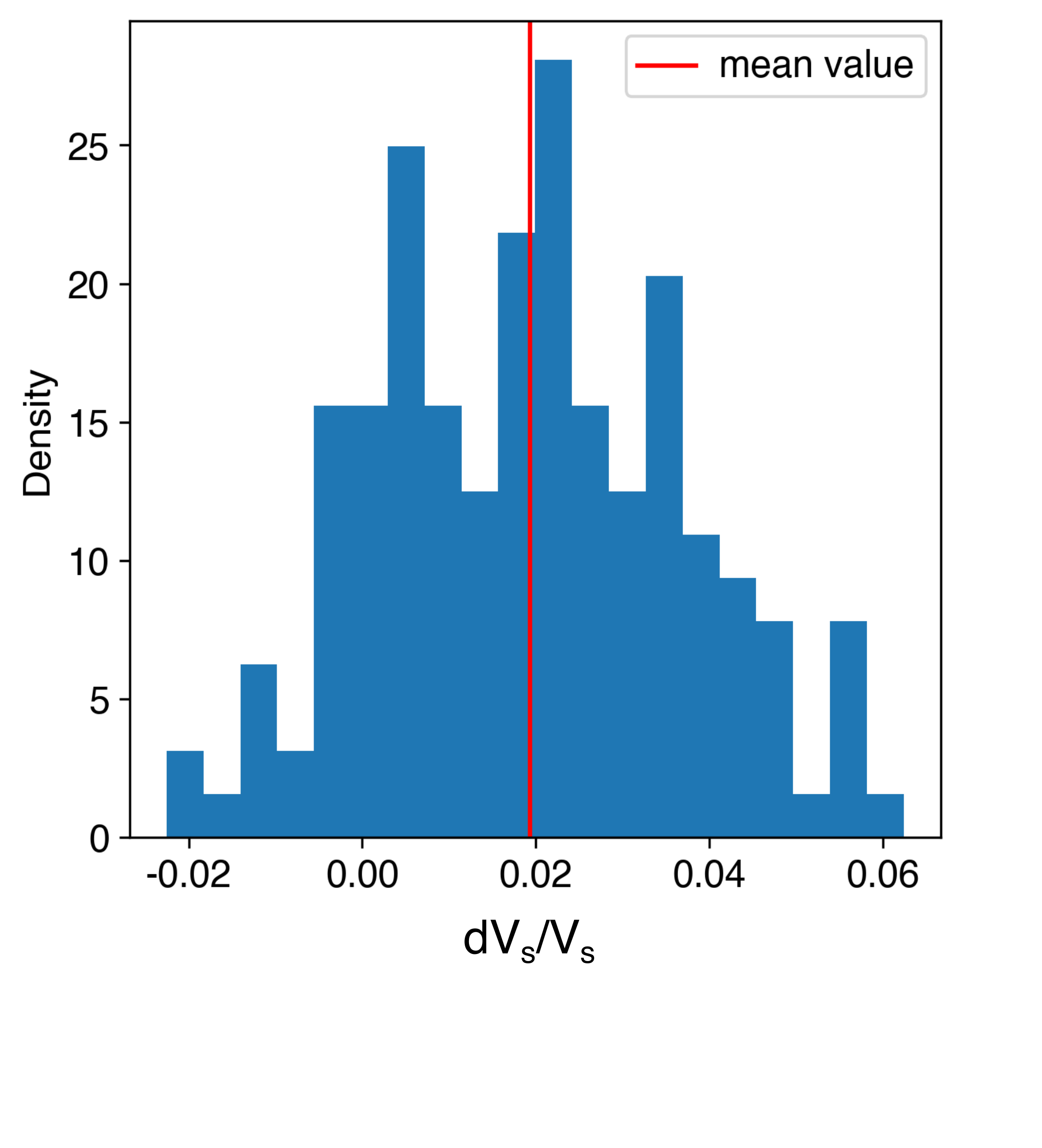}
  \centering
  \caption{\textbf{Relative volume change.} Distribution of relative volume change of assembled nanostructure. The mean volume change (red line) is 0.019, close to the stretch/elongation ratio ($\lambda$ = 0.02). In other words, the assembled nanostructure stretch mainly along the axial direction of fibers and the cross section of the cylindrical structure remains almost constant. \label{fig:S_deformation_ratio}}
  
\end{figure}

\begin{figure}[htbp]
  \includegraphics[width=0.8\linewidth]{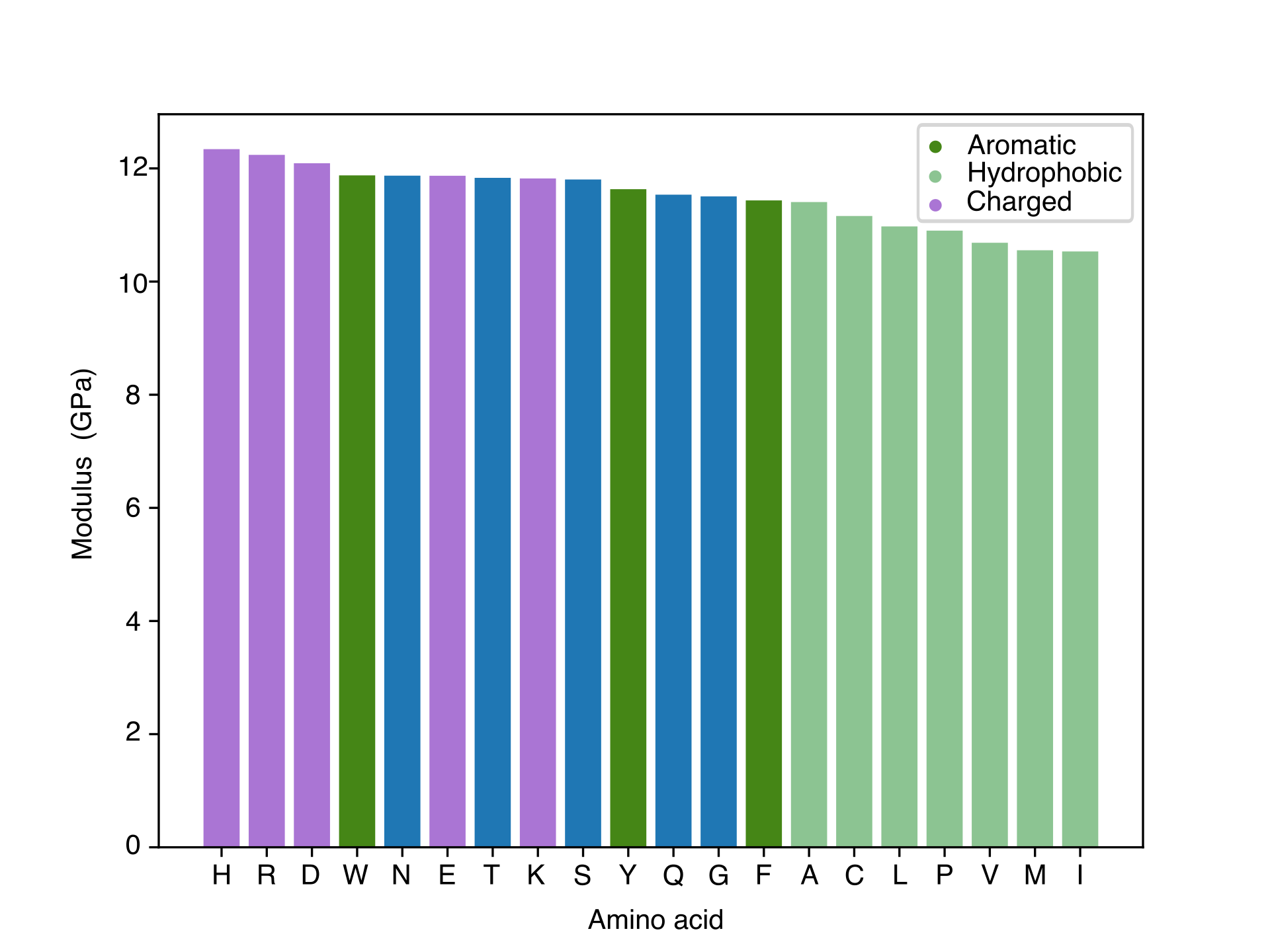}
  \caption{\textbf{Normalised modulus for each amino acid}. Modulus contribution from each amino acid.\label{fig:S_modulus_frequency}
  }
\end{figure}

\begin{figure}[htbp]
  \includegraphics[width=\linewidth]{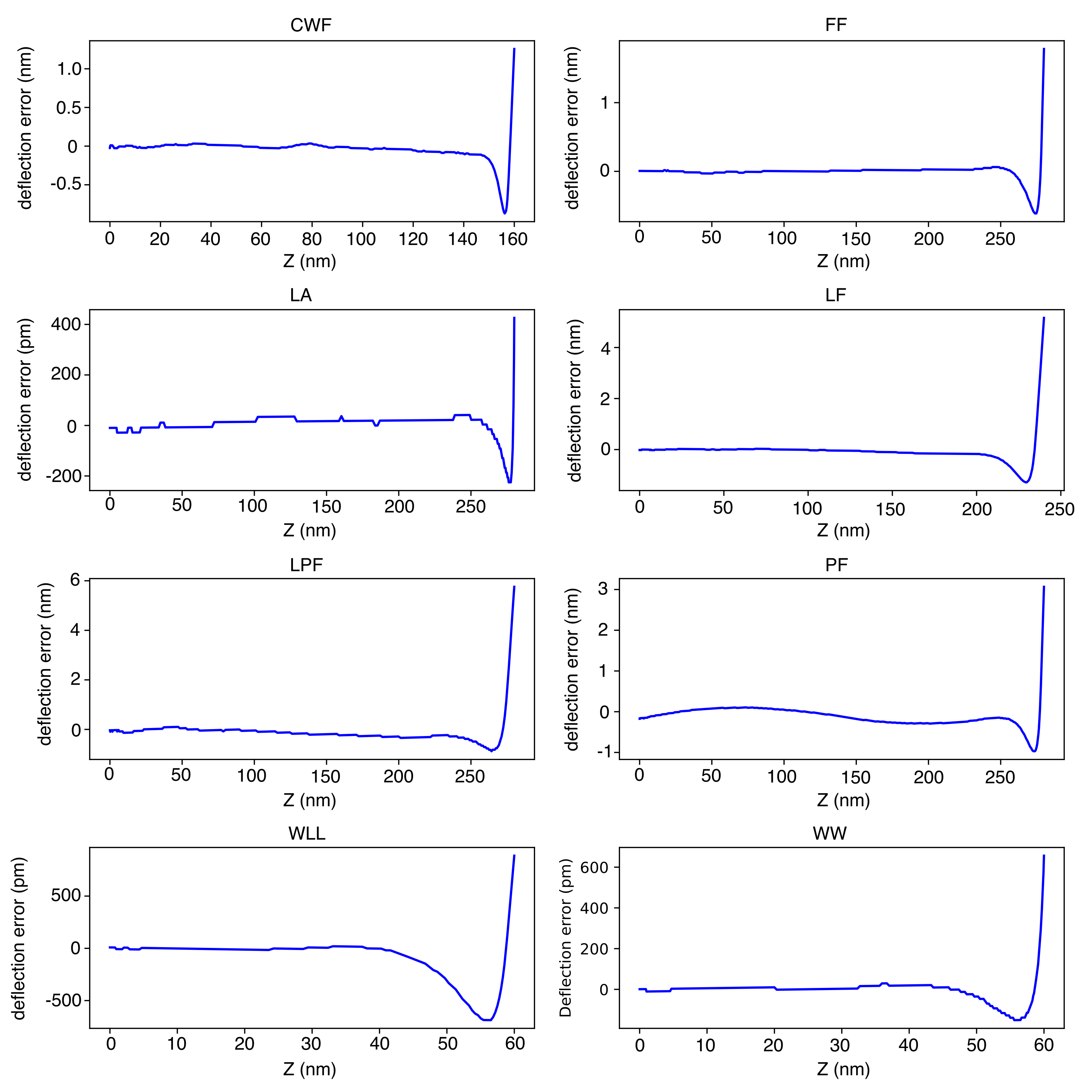}
  \caption{\textbf{Example force curve for each peptide}. Curves are measured from tip retraction, and examples of the collection of curves used to find to find the DMT modulus.\label{fig:retraction}}
  
\end{figure}
\clearpage

\begin{figure}[htbp]
  \includegraphics[width=\linewidth]{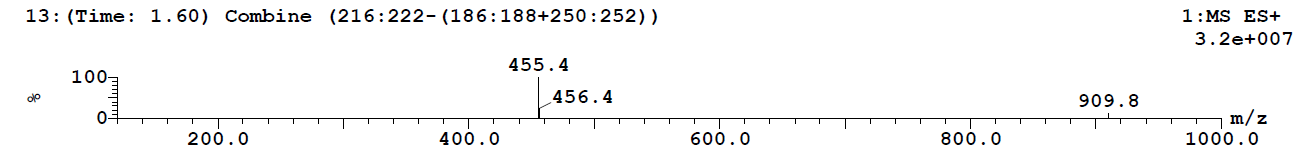}
  \caption{\textbf{Mass spectrum (ES+) from LCMS of CWF.}\label{fig:MS_CWF}}
  
\end{figure}

\begin{figure}[htbp]
  \includegraphics[width=\linewidth]{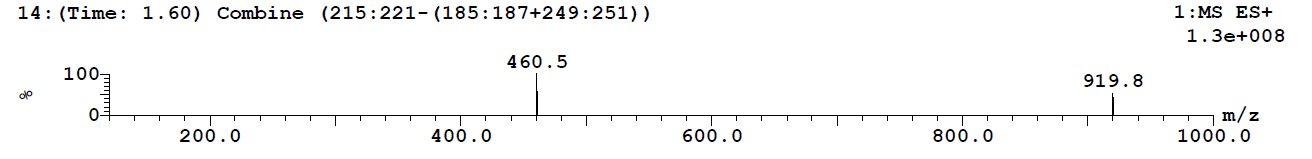}
  \caption{\textbf{Mass spectrum (ES+) from LCMS of FFF.}\label{fig:MS_FFF}}
  
\end{figure}

\begin{figure}[htbp]
  \includegraphics[width=\linewidth]{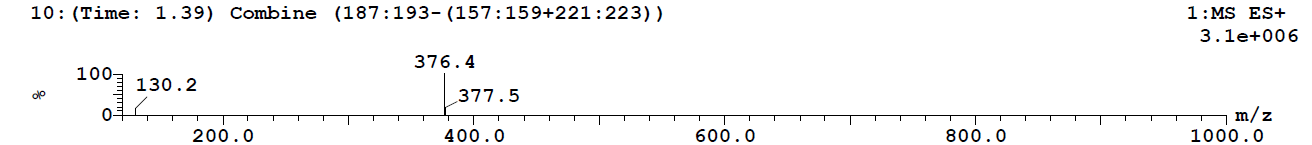}
  \caption{\textbf{Mass spectrum (ES+) from LCMS of LPF.}\label{fig:MS_LPF}}
  
\end{figure}

\begin{figure}[htbp!]
  \includegraphics[width=\linewidth]{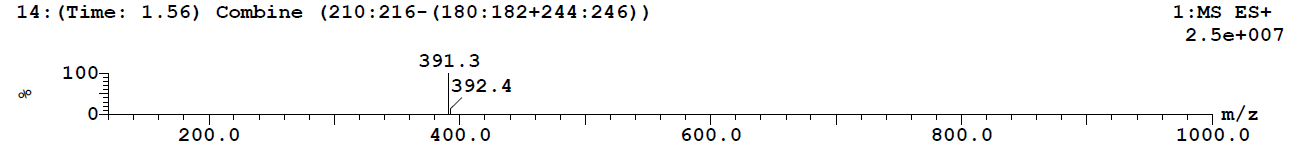}
  \caption{\textbf{Mass spectrum (ES+) from LCMS of WW.}\label{fig:MS_WW}}
  
\end{figure}

\begin{figure}[htbp]
  \includegraphics[width=\linewidth]{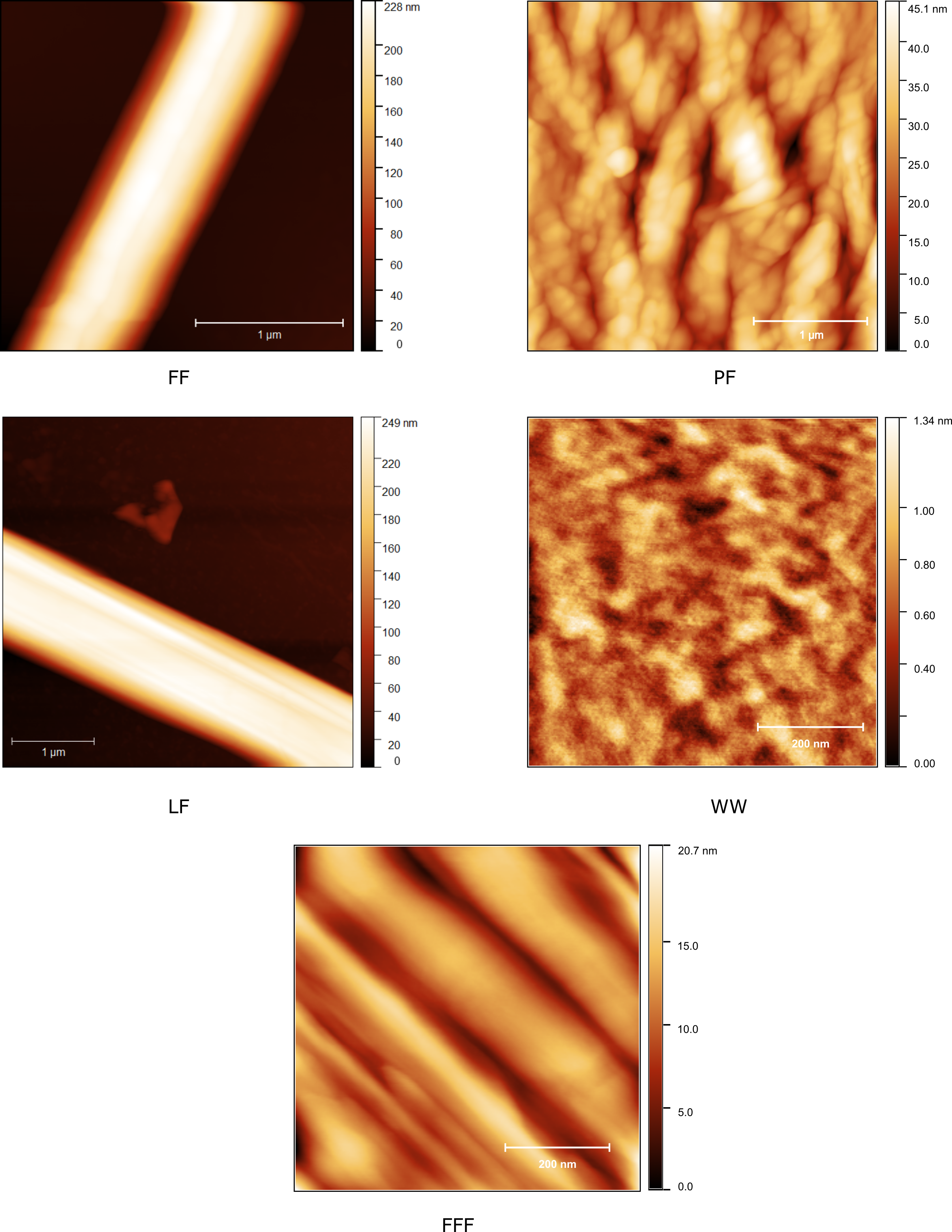}
  \caption{\textbf{AFM images for each peptide,} where not presented in the main text.\label{fig:AFM}}
  
\end{figure}
\end{document}